\documentclass[11pt,a4paper]{article}
\usepackage{jheppub}
\usepackage{graphicx}
\usepackage{amsmath}
\usepackage{amssymb}
\usepackage{amsthm}
\usepackage{empheq}
\usepackage{relsize}
\usepackage{epstopdf}
\usepackage{url}
\usepackage{hyperref}
\usepackage[footnotesize]{caption}
\theoremstyle{remark}

\theoremstyle{definition}

\theoremstyle{remark}

\begin{document}

  \title{\LARGE The Fate of Monsters in Anti-de Sitter Spacetime}
  \author[1,2]{\large Yen Chin Ong,}
  \author[1,2,3,4]{\large Pisin Chen}

 \affiliation[1]{Graduate Institute of Astrophysics, National Taiwan University, Taipei 10617, Taiwan.}
 \affiliation[2]{Leung Center for Cosmology and Particle Astrophysics,\\
 National Taiwan University, Taipei 10617, Taiwan.}
  \affiliation[3]{Department of Physics, National Taiwan University, Taipei 10617, Taiwan.}
 \affiliation[4]{Kavli Institute for Particle Astrophysics and Cosmology, SLAC National Accelerator Laboratory, Stanford University, Stanford, CA 94305, U.S.A.}

\emailAdd{ongyenchin@member.ams.org}
 \emailAdd{chen@slac.stanford.edu}

\abstract{
Black hole entropy remains a deep puzzle: where does such enormous amount of entropy come from? Curiously, there exist gravitational configurations that possess even larger entropy than a black hole of the same mass, in fact, \emph{arbitrarily high} entropy. These are the so-called \emph{monsters}, which are problematic to the Anti-de Sitter/Conformal Field Theory (AdS/CFT) correspondence paradigm since there is far insufficient degrees of freedom on the field theory side to account for the enormous entropy of monsters in AdS bulk. The physics of the bulk however may be considerably modified at semi-classical level due to the presence of branes. We show that this is especially so since monster spacetimes are unstable due to brane nucleation. As a consequence, it is not clear what the final fate of monsters is. We argue that \emph{in some cases} there is no real threat from monsters since although they are solutions to Einstein's Field Equations, they are very likely to be completely unstable when embedded in string theory, and thus probably are not solutions to the full quantum theory of gravity. Our analysis, while suggestive and supportive of the claim that such pathological objects are not allowed in the final theory, \emph{by itself} does not rule out \emph{all} monsters. We comment on various kin of monsters such as the ``bag-of-gold'' spacetime, and also discuss briefly the implications of our work to some puzzles related to black hole entropy.}

\maketitle
\flushbottom

\section{Introduction: Monsters and Their Kin}

The term ``monsters'' was coined by Stephen Hsu and David Reeb \cite{HR, HR2} to refer to pathological configurations that possess entropy greater than their area in Planck units. Monsters have finite ADM (Arnowitt-Deser-Misner) mass and surface area, but potentially unbounded entropy. The idea for constructing such configurations is not difficult: In flat space, given an area that bounds a volume, we have good knowledge about the volume of the interior. In particular, as the area shrinks, so does the volume; this is however not necessarily true in curved spaces, since one can have larger proper volume than expected from looking at the surface area alone. The idea is not new, for example, Wheeler's ``bag-of-gold'' spacetime \cite{Wheeler} (closed FRW universe glued across Einstein-Rosen Bridge) is such an example; see also discussion in \cite{SWJ}. In other words, curved space can hide large amount of ``stuffs''.

As it turns out, monsters are problematic with regard to both unitary evaporation of black holes and Anti-de Sitter/Conformal Field Theory (AdS/CFT) correspondence. Since Einstein's General Theory of Relativity allows for such solutions, if we argue that monsters somehow cannot arise, then the prevention mechanism must lie in the realm of \emph{quantum} theory. In Section 2, with the hope that this work is self-contained, we will briefly review the construction of a monster and recall why such pathological configurations are problematic for AdS/CFT correspondence, as well as some ideas that have been proposed to banish monsters and their kin from quantum gravity. In Section 3, we introduce Seiberg-Witten instability for asymptotically locally hyperbolic Riemannian manifold, which we then employ to investigate the stability of AdS monsters in Section 4. We conclude with discussions about various kin of monsters, and puzzles regarding black hole entropy.  

The main idea to keep in mind is as follows: In string theory, asymptotically locally AdS geometry becomes unstable if certain brane action becomes negative. If the negative action is not bounded below, then such configurations are inherently unstable, and if furthermore no finite operations can evolve said configurations to new ones with non-negative brane action, then this can be interpreted as the full theory not admitting such pathological configurations. Using such argument, we are able to slay some, but not all, monsters. This nevertheless means that in all cases in which our method is applicable, monsters, even if they exist, cannot have \emph{arbitrarily} large entropy, i.e. Seiberg-Witten instability naturally allows us to bound the size of monsters. Unfortunately, with Seiberg-Witten instability as the only tool to investigate the existence of monsters, one finds that not all monsters and their various kin can be addressed in this approach. In particular, our argument works well only for the cases in which the region that holds large entropy is \emph{not} behind a black hole horizon -- a den for monsters to hide and avoid being slain. This means that it leaves open the interesting issue of whether the ``bag-of-gold'' spacetime, where the FRW universe is interior to the Schwarzschild event horizon, is consistent with the full quantum theory of gravity.   

\section{A Monster is Born}

As detailed by Hsu and Reeb \cite{HR, HR2}, monsters configurations are expected to quickly undergo gravitational collapse to form black hole. Consider, for example, a spherically symmetric initial data on a Cauchy slice $\Sigma_0$ at a moment of time symmetry, which is \emph{not yet a black hole}, i.e., there is no marginally trapped surface. Intuitively ``moment of time symmetry'' means that the configuration is initially ``at rest''. Mathematically, an ``instant of time'' is described by a spacelike hypersurface $\Sigma$, which is a Riemannian manifold with metric $h_{ij}$. The initial data must satisfy the \emph{initial value constraints} determined by Einstein's contraint equations (See e.g. Chapter 10 of \cite{Wald})
\begin{subequations}
\begin{empheq}[left=\empheqlbrace]{align}
D^i \left(K_{ij} - K^l_{~l}h_{ij}\right) &= -8\pi \mathcal{J}_{j}, \label{1a} \\
R(h) + (K^i_{~i})^2 - K_{ij}K^{ijj} &= 16\pi \rho,
 \end{empheq}
\end{subequations}
where $D$ is the covariant derivative operator in $\Sigma$, $R(h)$ is the scalar curvature of $\Sigma$ defined by $h_{ij}$, $K_{ij}$ is the extrinsic curvature of $\Sigma$, $\rho$ is the value of the energy density of the matter fields on $\Sigma$ as measured by observers whose worldlines are perpendicular to $\Sigma$, and finally $\mathcal{J}^j$ is the projection onto $\Sigma$ the four-dimensional energy-momentum flux vector seen by the observers. By ``$\Sigma_0$ is at a moment of time symmetry'' one means that the extrinsic curvature, $K_{ij}$, of the hypersurface representing that instant of time vanishes. By Eq.(\ref{1a}), this implies that $\mathcal{J}^j$ must be zero, and thus the four-dimensional energy-momentum flux vector must be orthogonal to the initial hypersurface $\Sigma_0$, hence ``symmetric'' with respect to $\Sigma_0$. 

The spherical symmetric configuration has spatial metric 
\begin{equation}\label{metric}
\left. ds^2\right|_{\Sigma_0} = g_{rr}dr^2 + r^2d\Omega^2,
\end{equation}
where $d\Omega^2$ is the standard metric on a 2-sphere.
Since the configuration is not yet a black hole, the full spacetime metric takes the form
\begin{equation}
ds^2=-g_{tt}dt^2 + g_{rr}dr^2 + r^2 d\Omega^2,
\end{equation}
where $g_{tt}$ and $g_{rr}$ are not necessarily static, that is they can be functions of $t$ in addition to $r$. Typically, one considers a ``star'' of certain fluid with some kind of density profile $\rho$ such that $g_{tt}$ and $g_{rr}$ take the same form as black hole metric \emph{exterior} to the star, however it is important to note that unlike black holes, $g_{tt}g_{rr}\neq -1$ in the case of fluid \cite{Ted}. For asymptotically flat case, it is well known that Einstein's field equations determine 
\begin{equation}\label{grr}
g_{rr}^{~-1} = 1- \frac{2M(r)}{r},
\end{equation} 
where 
\begin{equation}
M(r) = 4\pi \int_0^r r'^2 \rho(r') dr'
\end{equation}
is the ``energy within radius $r$''. Note that $\rho(r)=\rho(r,t_0)$ is the \emph{proper} energy density as seen by a stationary observer at $r$ on the initial time slice $t=t_0$.

Note that the ``mass'' $M(r)$ is defined via integrating over the \emph{flat space} volume element $r^2\sin\theta dr \wedge d\theta \wedge d\phi$, so it is \emph{not} the \emph{proper} mass. The latter is defined by
\begin{equation}
M_p(r) = 4\pi \int_0^r r'^2 \sqrt{g_{rr}} \rho(r') dr'.
\end{equation}   
From the Schwarzschild solution (valid as exterior solution), we know that $M=\lim_{r\to\infty}M(r)$ is the ADM mass, which can be interpreted as the \emph{total} energy of the spacetime, including gravitational energy. However $\rho(r)$ is the density profile of the star and this does \emph{not} include gravitational energy. The difference between $M$ and $M_p$ is precisely the negative gravitational binding energy. Monsters can thus be viewed as configurations whose otherwise huge mass is canceled by the negative gravitational binding energy. 

The same story holds in asymptotically anti-de Sitter spacetimes, but now 
\begin{equation}\
g_{rr}^{~-1} = 1- \frac{2M(r)}{r} + \frac{r^2}{L^2},
\end{equation} 
where $L$ is the length scale associated with the cosmological constant by $\Lambda=-3/L^2$ \cite{Stuchlik}. It is convenient to denote 
\begin{equation}
\epsilon(r) = g_{rr}^{~-1}.
\end{equation}

Hsu and Reeb \cite{HR} gave two examples of monster configurations in asymptotically flat spacetime: the blob and the shell. The blob refers to an object with core radius $r_0$ and mass $M_0$ and density profile
\begin{equation}
\rho(r)=\rho_0\left(\frac{r_0}{r}\right)^2, ~~ r_0 < r < R.
\end{equation} 
In the unit $8\pi \rho_0 r_0^2 = 1$, we have
\begin{equation}
\epsilon(r) = \epsilon_0\left(\frac{r_0}{r}\right),
\end{equation}
where $\epsilon_0 = 1-2M_0/r_0$. Hsu and Reeb showed that the entropy of the blob monster is
\begin{equation}
S \sim \frac{\rho^{\frac{3}{4}}r_0}{\sqrt{\epsilon_0}}R^2.
\end{equation}
The entropy can thus be arbitrarily large by taking $\epsilon_0$ arbitrarily small. Furthermore, we can obtain faster than area scaling by taking $\epsilon(r)$ to approach zero faster than $1/r$. More generally \cite{HR2}, one can consider profile of the power-law type:
\begin{equation}
\epsilon(r) = \epsilon_0\left(\frac{r_0}{r}\right)^{\gamma}, ~~\gamma > 0.
\end{equation}

The shell monster, on the other hand, is constructed from a thin shell of material with $R < r < R+d$ such that the mass function is
\begin{equation}\label{shell}
M(r) = \begin{cases}
\displaystyle \frac{R_1 (r-R_0)(1-\epsilon(r))}{2(R_1-R_0)}, ~~\text{if}~ R_0<r<R_1,\\
\\
\displaystyle \frac{r(1-\epsilon_0)}{2}, ~~\text{if}~ R_1 < r < R_1+d=R.
\end{cases}
\end{equation}
Here the function $\epsilon(r)$ decreases rapidly to some $\epsilon_0$ betwee $R_0<r<R_1$, and is constant $\epsilon(r)=\epsilon_0$ for $R_1 < r < R$.
Again, by choosing arbitrarily small $\epsilon_0$, the entropy in the region $R_1 < r < R$ can be made arbitrarily large. 

The major problem posed by monsters and their kin is the following: monsters are believed to inevitably evolve into black holes \cite{HR, HR2} (although we will later argue \emph{against} this, at least in the case of some monsters in anti-de Sitter spacetime); however by construction, the entropy on the initial Cauchy slice can be arbitrarily larger than the entropy of the eventual black hole, where the latter is determined by the area of the event horizon according to Bekenstein-Hawking formula $S = A/4$. As the black hole evaporates, the entropy released is only at the order of $M^2$. Assuming the usual scenario in which the black hole completely evaporates, in order to preserve unitarity one would need to remove monsters with $S \gtrsim A$ from the associated Hilbert space. In the case where one considers black hole remnant instead of complete evaporation, as long as the end state is not a remnant that locks up enormous amount of entropy (see, e.g. \cite{ACS} in which the entropy of black hole remnant remains small), the same puzzle remains (this of course depends on how one \emph{interprets} the Bekenstein-Hawking entropy. We leave this issue to the discussion section). Likewise, in the contexts of AdS/CFT correspondence, monsters are problematic because on the gravity side we have enormous amount of entropy but there is far from enough degrees of freedom on the field theory side to describe such a configuration. Thus, as pointed out already in \cite{HR}, monsters with sufficiently high entropy are semiclassical configurations with \emph{no} corresponding microstates in a quantum theory of gravity. In view of increased evidence in support for AdS/CFT correspondence, it is desirable to understand the nature of monster spacetime in this context. We thus focus our attention to the fate of monsters in AdS, since gravity in the bulk may be considerably modified at semi-classical level by the existence of extended objects such as D-branes. 

We remark that it has been argued that the current known laws of physics prevent creation of monster configurations even by arbitrarily advanced civilization \cite{HR, SWJ}. Likewise, at the classical level no mechanism is known to create bag-of-gold spacetime from empty AdS space by acting with boundary observables \cite{Marolf}. However, there seems to be no obvious reasons monsters and its kin cannot be created via some quantum tunneling processes \cite{HR, HR2, Marolf}.

There are at least two ways out of this puzzle:

\begin{itemize}
\item[(1)] There might exist superselection rule that prevents the formation of monsters and its kin from quantum tunneling processes. However there is no obvious reason why this should be the case. Marolf suggested in the context of bag-of-gold spacetime in AdS that since such spacetime inevitably contains a past singularity, there is no obvious way to construct a bag-of-gold spacetime by simply manipulating perturbative excitations near the AdS boundary, and this could be a hint that bag-of-gold spacetime lies in different superselection sector of the theory \cite{Marolf} . This is also true for monsters: They evolve into black holes, but their \emph{time reversed evolution} also leads to black hole formation, that is, in the time-forward sense, monsters emerge out of a \emph{white hole} singularity in the past \cite{HR, HR2}. This is due to the fact that by construction the initial data set is time symmetric.

\item[(2)] A full theory of quantum gravity might not permit monsters and its kin to exist. For example, Hsu and Reeb \cite{HR3} showed that, assuming unitarity and no remnant and no topology change, there must exist a one-to-one correspondence between states on future null and timelike infinity and on any earlier spacelike Cauchy surface. Consequently a large set of semiclassical spacetime configurations including the monsters and bag-of-gold spacetime are \emph{excluded} from the Hilbert space of quantum gravity. Presumably if the end state of black hole evaporation is a remnant, but one which does not lock up very high entropy, it is conceivable that pathological configurations with sufficiently high entropy would still be excluded from the Hilbert space.  
\end{itemize}

In support of the claim that monsters simply do not exist in quantum gravity, we will argue via Seiberg-Witten instability that configurations with finite area bounding sufficiently large volume is not acceptable in string theory. This idea has been previously pointed out by McInnes \cite{McInnes} in the context of ``Bubble de Sitter with Casimir effect'' spacetime, which has precisely the aforementioned property that finite area bounds large volume at sufficiently large value of proper time. In this work we will emphasize this idea again, and explicitly show that in particular the shell monster are completely unstable in string theory (the case for blob monster is similar). Furthermore, as pointed out also in \cite{McInnes}, many concrete examples of Seiberg-Witten instabilities are induced by violation of the Null Energy Condition (NEC). See also \cite{NEC} for general discussion about how violation of NEC implies instability in a broad class of physical models. We will prove that the shell monsters indeed also violate the NEC. Nevertheless, we will point out how this does not rule out \emph{all} monsters. 

\section{Seiberg-Witten Instability}

We shall now review the notion of Seiberg-Witten instability. For any locally asymptotically AdS spacetime in Type IIB string theory defined on $\text{AdS}_{n+1} \times S^{9-n}$, $n \geq 3$, the \emph{Seiberg-Witten brane action} can be defined on the Euclidean signature spacetime obtained after performing Wick rotation by 
\begin{equation}\label{SWaction}
S=\Theta (\text{Brane Area}) - \mu (\text{Volume Enclosed by Brane}).
\end{equation}
Here $\Theta$ is related to the tension of the brane and $\mu$ relates to the charge enclosed by the brane due to the background antisymmetric tensor field.
We remark that the type IIB backgrounds
assumed here are of Freund-Rubin type \cite{FR}, i.e. the $\text{AdS}_{n+1} \times S^{9-n}$ spacetime metric is supported by the background antisymmetric field strength. In string theory, the existence of such flux field naturally induces compactification so that the full 10-dimensional spacetime reduces to a product manifold $\text{AdS}_{n+1} \times S^{9-n}$ where the factor $S^{9-n}$ is compactified.

We see that the action will become negative if the term proportional to volume is large. The most dangerous situation occurs when the charge $\mu$ attains its maximal value: the BPS case with $\mu=n\Theta$. Explicitly, the Seiberg-Witten brane action is given by
\begin{equation}
S[r]= \Theta ~r^{n-2} \int d\tau \sqrt{g_{\tau\tau}} \int d\Omega  - n\Theta \int d\tau \int^r dr' \int d\Omega ~r'^{n-1}\sqrt{g_{\tau \tau}}\sqrt{g_{r'r'}},
\end{equation} 
where $g_{\tau\tau}=-g_{tt}$.

This action describes a probe brane that we introduce to investigate the background fields and geometry of the bulk. Seiberg and Witten showed that non-perturbative instability occurs when the action becomes negative due to uncontrolled brane productions \cite{SeibergWitten, Kleban}. Brane-anti-brane pairs are always spontaneously created from the AdS vacuum, a phenomenon analogous to the well-known Schwinger effect in quantum electrodynamics \cite{Schwinger}, with the rate of brane-anti-brane pair production being proportional to $\exp(-S)$ where $S$ is the Seiberg-Witten brane action. If the action becomes negative, the AdS vacuum will nucleate brane-anti-brane pairs at exponentially large rate instead of exponentially suppressed. This disrupts the background geometry so much so that the spacetime is no longer described by the metric that we started with. That is to say, the original spacetime is not stable if such brane-anti-brane production is exponentially enhanced due to the ``reservoir of negative action''. Seiberg-Witten instability can occur, e.g. if the Seiberg-Witten brane action is negative at large $r$ limit, which can happen when the boundary has negative scalar curvature \cite{SeibergWitten}. To understand Seiberg-Witten instability in terms of brane and anti-brane dynamics in Lorentzian picture, see e.g. \cite{Barbon}. 

Seiberg-Witten instability applies to any spacetime of dimension $d=n+1 \geq 4$ with an asymptotically hyperbolic Euclidean version (loosely speaking, we say that a Riemannian manifold is asymptotically hyperbolic if it has a well-defined conformal boundary; this can of course be made precise), even for string theory on $X^{n+1} \times Y^{9-n}$, where $X^{n+1}$ is an $(n+1)$-dimensional non-compact asymptotically hyperbolic manifold (generalizing $\text{AdS}_{n+1}$) and $Y^{9-n}$ is a compact manifold (generalizing $S^{n+1}$)\footnote{Again, we remind the readers that such product manifold is supported by Freund-Rubin type of background antisymmetric tensor field. The requirement that $X^{n+1}$ has well-defined conformal boundary guarantees its volume form $\omega$ is exact: $\omega=d\mathcal{H}$ for an $n$-form $\mathcal{H}$, and $d\mathcal{H}$ is precisely the background antisymmetric tensor field of the appropriate supergravity theory on $X^{n+1}$.} \cite{McInnes5}. Furthermore, since the idea of holography and AdS/CFT correspondence is expected to arise in \emph{any consistent theory} of quantum gravity, we should expect that extended objects like branes to be generic in any consistent theory of quantum gravity, and also that phenomenon qualitatively similar to Seiberg-Witten instability to continue to be present (See also Footnote (1) in \cite{BKLS}). 

Our plan is to show that monsters are unstable objects in the Seiberg-Witten sense. However, it is crucial to note that \emph{merely being unstable is not good enough} to rule out monsters. After all, physical states which are unstable usually do not exhibit truly runaway behavior. As argued, the brane-anti-brane pairs will soon occupy the surrounding black hole environment due to the exponential rate of pair-production. This will likely alter the boundary conditions of the original action. As a result, the exponential pair-production will stop. We can compare this with the more familiar physical system of neutral hydrogen gas ionizing into plasma (which we can think of as ``pair-production'' of electrons and protons) under external $E$-field between parallel plates. As the $E$-field reaches the atom's ionization energy within one \AA, there will be an exponential avalanche. This will nevertheless be quickly suppressed since the surrounding plasma would induce negative $E$-field that will counter the original $E$-field. In other words, physical instability is often \emph{self-limiting} due to backreaction. We might therefore object that even if monsters are \emph{unstable} objects in string theory, they can still \emph{exist} for at least short amount of time before evolving into other stable configurations (perhaps a black hole). This clearly does not solve the problem since monster states \emph{are expected to evolve into black holes in the first place}, i.e. the problem persists since however brief the life span of a monster is due to instability, we cannot account for that existence on the field theory side of the AdS/CFT correspondence. Therefore, it is important to argue that not only monsters are unstable, but  also that they are \emph{completely unstable} in the sense that no backreaction can bring the configuration to \emph{any} stable configuration, and therefore monsters probably do not exist in the full theory of quantum gravity. 

For example, if the geometry of (Euclidean) locally hyperbolic spacetime is such that its scalar curvature at the boundary is negative, and it is of certain class of topology, then using results from differential geometry, one can prove that no matter how the branes deform the spacetime, the scalar curvature at infinity can never become everywhere positive or zero \cite{McInnes5}. At this stage, this suggests that we can already eliminate an entire zoo of locally AdS monsters spacetime with negative curvature at infinity. We nevertheless need to be careful since we need to take into account the \emph{time scale} for the instability to set in. We will come back to this point shortly. One might also argue that \emph{infinite} reservoir of negative action means that the action cannot be made everywhere positive by pair-production of \emph{finitely} many brane-anti-brane pairs. This innocent-sounding reasoning, however, is a non-trivial statement since as we shall explain, it is not true in general. Despite that, as we will subsequently argue, such reasoning is still applicable for these monsters.  

What if the brane action is only negative for some finite range of $r$, and so the reservoir of negative action is finite? For such action, the emitted branes and antibranes can minimize the action by moving to this region (Note that most of the brane-anti-brane pairs are actually created in such region in the first place due to the exponential enhancement in pair-production rate, and by causality if nothing else) instead of collapsing to zero size under their own tension. Nevertheless the action can only be reduced by a \emph{finite} amount in this case. This leads Maldacena and Maoz \cite{MaldacenaMaoz} to suspect that there should be ``nearby'' solutions that are stable. In a more dynamical picture, the branes are  produced in such a way that some of the metric parameters of the black hole spacetime will eventually be brought down below the threshold value that triggered the instability. However, when everything has settled down to a stable configuration, it is \textit{no longer the original spacetime}. It has become a ``nearby'' solution in the sense of Maldacena and Maoz.  \emph{Qualitatively} we will expect that the more ``negative reservoir'' the action has, the more unstable it is, in the sense that ``nearby'' solutions may not even exist and so the spacetime is completely unstable (and consequently is likely to \emph{not} be a solution of the full theory), although we do not yet have a quantitative treatment of this claim. We will see that there exist, in fact, monsters of this class.

\section{Monsters in String Theory}

It was commented in \cite{Marolf} that ``AdS/CFT appears to predict that such tunneling (to create bag-of-gold spacetime) is not possible and that understanding this prediction from the AdS gravity point of view remains an important open problem''. We argue that the solution to the analogous problem regarding (at least some) monsters lies in Seiberg-Witten instability. The reason we specifically emphasize monsters separately from the bag-of-gold spacetime will become clear later. Specifically, we consider any spherically symmetric monster configurations with metric given by Eq.(\ref{metric}), with $g_{tt}$ and $g_{rr}$ attaining the functional form of their vacuum black hole counterpart in the exterior spacetime. This is a ``star'' with unusual density profile. Since monsters are very close to being a black hole ($\epsilon \approx 0$) but \emph{still not} a black hole, $r=r_{\text{eh}}$ does not represent a horizon but a surface inside the star. For convenience, we denote $f \equiv g_{tt}$ and $\epsilon \equiv g_{rr}^{~-1}$. Despite the fact that monsters are unstable and therefore dynamical objects, we will take the metric to be static just to study the qualitative behavior of monsters spacetime (Essentially we are treating the case in which the dynamical spacetime can be approximated by a sequence of static spacetimes, i.e. quasi-static approximation).  

We can then compute the Seiberg-Witten brane action:
\begin{equation}\label{action}
S \propto r^2f^{\frac{1}{2}} - \frac{3}{L}\int_{r_{\text{eh}}}^r r'^2\left(f\epsilon^{-1}\right)^{\frac{1}{2}} dr', 
\end{equation}
where we have omitted an overall \emph{positive} multiplicative factor.
It is already suggestive at this point that the second term is going to dominate if $\epsilon$ is sufficiently small, which then leads to negative brane action. We must however be more careful and explicit about the details. First of all, we remark that Eq.(\ref{action}) is \emph{not} exactly correct. In the case of black holes, when we do a Wick-rotation to Euclidean field theory, \emph{the black hole interior is removed}, and thus it makes sense that the radial integral in Eq.(\ref{action}) starts from the horizon $r=r_{\text{eh}}$. For a spacetime without horizon, like our fluid star, strictly speaking the integral should start from $r=0$. But in writing Eq.(\ref{action}), all we want is to do a fair comparison, since the only thing we can compare between a star and a black hole of the same mass is their exterior geometries. Here by ``exterior'' we mean outside of the event horizon, which for the star lies in its \emph{interior} (in the formal sense, since it is not a ``real'' horizon). Regardless, the geometries outside the event horizon for both a star and a black hole is not too different and thus can be compared. More crucially in our work below, the objective is to show that the brane action can become negative, and since the inclusion of the range $[0,r_{\text{eh}}]$ \emph{only makes the action more negative}, for simplicity we need not consider said range in our radial integral. Similarly, we can assume that the horizon $r_{\text{eh}}$ lies within the ``problematic region''  $R_1\leq r \leq R$ (i.e. region in which, by construction, arbitrarily high entropy can be contained). However this is due to the fact that, in the construction of the shell monster, our problematic region shares boundary with the exterior vacuum geometry, and so we at least know that its horizon \emph{lies inside} the boundary surface $r=R$. In general, one can imagine constructing a monster configuration such that the problematic region lies strictly within a fluid, and it is then possible that the horizon is exterior to the the region, i.e. $r_{\text{eh}} > R$. In such case of ``concentric monster'' consists of problematic region bounded between relatively normal fluid, our calculation below needs to be modified accordingly, and then it is crucial that our range of integration be taken from origin instead of the horizon.
For the shell monster with mass profile given by Eq.(\ref{shell}), the brane action is
\begin{equation}\label{shellaction}
S[\text{shell}] \propto r^2f^{\frac{1}{2}} - \epsilon_0^{-\frac{1}{2}}\frac{3}{L}\int_{r_{\text{eh}}}^R r'^2 f^{\frac{1}{2}} dr' - \frac{3}{L}\int_{R}^r r'^2\left(f\epsilon^{-1}\right)^{\frac{1}{2}} dr', 
\end{equation}
the second term becomes unbounded below if $\epsilon_0$ is arbitrarily small. The case for the blob monster is similar. We should of course be more careful about the first term and the last term, since the asymptotic values of their sum could very well be \emph{infinite}. We will discuss this in more details below. 

Let us first remark on the energy condition, since pathological constructions like the monsters immediately raise the concern about violation of energy condition. This is indeed true, at least for some monsters. To see this, as noted in \cite{McInnes2}, we can take the radial null vectors to be of the form
\begin{equation}
n^\mu = (\epsilon^{-\frac{1}{2}},\pm f^{\frac{1}{2}},0,0)^\text{T},
\end{equation}
so that consequently the Ricci tensor satisfies \cite{Ted}
\begin{equation}\label{NRC}
R_{\mu\nu}n^{\mu}n^{\nu} = \frac{(f\epsilon^{-1})'\epsilon}{r},
\end{equation}
where prime denotes derivative with respect to $r$, and superscript $T$ denote vector transposition.  

Typically in classical general relativity, various energy conditions are imposed on the matter field. The weakest of these energy conditions is the Null Energy Condition (NEC), which requires that the energy-momentum tensor $T_{\mu\nu}$ satisfies $T_{\mu\nu}n^{\mu}n^{\nu} \geq 0$ for all null vectors $n^\mu$. By the Einstein's Field Equations, the NEC is equivalent to the \emph{Null Ricci Condition} $R_{\mu\nu}n^\mu n^\nu \geq 0$. Consequently, if the NEC holds, we will have, by Eq.(\ref{NRC}), 
\begin{equation}
(f\epsilon^{-1})' \geq 0. 
\end{equation}
By virtue of AdS version of Birkhoff's theorem, in the exterior of our fluid star the metric coincides with that of a black hole, of which $f\epsilon^{-1}$ is precisely unity. Since $f\epsilon^{-1}$ is an increasing function of $r$, its value for NEC-nonviolating fluid is always \emph{less} than unity. For stable black holes (of vacuum Einstein solution, i.e. $k=0,1$ topological black holes, but \emph{not} $k=-1$ case, which is unstable in the Seiberg-Witten sense \cite{McInnes2}), the Seiberg-Witten brane action in Eq.(\ref{action}) is always positive, so for stable fluid, the action is also positive (since the second term, with the same domain of integration, is smaller than the black hole case). Consequently, \emph{if the NEC holds, then the brane action of the (exterior geometry of the) ``star'' is always positive}. The contrapositive statement is then: If the star has negative brane action, then the constituent fluid violates NEC. In other words, \emph{monsters, at least the blob and the shell species, violate NEC}. The argument here may not work for concentric monster since the exterior geometry does not contain any problematic region, and so as we discussed above, in considering Seiberg-Witten instability we should take the domain of integration to start from the interior of the horizon. Since the domain of integration is no longer the same as that of the black hole case, we cannot establish the argument above that leads to the conclusion of NEC violation. This is of course not saying that such monsters are free of energy condition violation. 

Now we return to the brane action for shell monster in Eq.(\ref{shellaction}). Imagine a ``toral star'' that collapses to form a black hole (for a careful treatment of such scenario, see \cite{SmithMann}) with flat horizon equipped with metric
\begin{equation}
ds^2 = -f(r)dt^2+f(r)^{~-1}dt^2 + r^2 d\Omega^2_{k=0},
\end{equation} 
where 
\begin{equation}
f(r)=\left(\frac{r^2}{L^2}-\frac{2\bar{M}}{r}\right)^{-1}, ~\bar{M}=\frac{M}{\pi K},
\end{equation}
and $d\Omega^2_{k=0}$ is the metric on compact flat submanifold, while $\bar{M}$ is the ADM mass $M$ divided by $\pi K$, where the parameter $K$ is related to the size of the flat compact manifold. Specifically, think of a flat square torus as parametrized by two circles of circumference $2\pi K$, and its area is $\Gamma_0 (T^2) = 4\pi^2 K^2$.  This equation then \textit{defines} $K$, so that for any compact flat 2-manifold, $K$ becomes a measure of the overall relative size of the space compared to the square torus. This follows the convention in \cite{McInnes2}. 

The brane action for such flat (uncharged) black hole asymptotes to a \emph{finite} positive constant \cite{McInnes2} 
\begin{equation}
\frac{\pi L \Theta A}{2},
\end{equation}  
where $\Theta$ is the brane tension and $A$ is the horizon area. For an ``ordinary'' toral star, i.e. when NEC is satisfied by the fluid, we observe that 
\begin{equation}
S[\text{ordinary star}] \propto r^2f^{\frac{1}{2}} - \frac{3}{L}\int_{r_{\text{eh}}}^r r'^2\left(f\epsilon^{-1}\right)^{\frac{1}{2}} dr'. 
\end{equation}
By Birkhoff's theorem, the exterior metric should match that of the black hole (i.e., $f\epsilon^{-1}=1$ for $r> R$), and so the brane action for monster $S[\text{Monster}(k=0)](r)$, should satisfy
\begin{flalign}
S[\text{Monster}(k=0)](r) &\propto r^2f^{\frac{1}{2}} - \frac{3}{L}\int_{r_{\text{eh}}}^R r'^2\left(f\epsilon^{-1}\right)^{\frac{1}{2}} dr' - \frac{3}{L}\int_{R}^r r'^2\left(f\epsilon^{-1}\right)^{\frac{1}{2}} dr' \\
&=   r^2f^{\frac{1}{2}} - \frac{3}{L}\int_{r_{\text{eh}}}^R r'^2\left(f\epsilon^{-1}\right)^{\frac{1}{2}} dr' - \frac{3}{L}\int_{R}^r r'^2 dr' \label{neg}\\
&= S[\text{Black Hole}(k=0)](r) + \frac{3}{L}\int_{r_{\text{eh}}}^R \left[r'^2\left[1-\left(f\epsilon^{-1}\right)^{\frac{1}{2}}\right] dr'\right].  \label{1} 
\end{flalign}
Taking limit $r \to \infty$ on both sides of the equation yields
\begin{equation}
\lim_{r \to \infty} S[\text{Monster}(k=0)](r) \propto
 \frac{\pi L \Theta A}{2} + \frac{3}{L}\int_{r_{\text{eh}}}^R \left[r'^2\left[1-\left(f\epsilon^{-1}\right)^{\frac{1}{2}}\right] dr'\right] \label{starboundary}.
\end{equation}

We note that even for $k=\pm 1$, the same argument in Eq.(\ref{1}) also yields 
\begin{equation}\label{shift}
S[\text{Monster}](r) = S[\text{Black Hole}](r) + \frac{3}{L}\int_{r_{\text{eh}}}^R \left[r'^2\left[1-\left(f\epsilon^{-1}\right)^{\frac{1}{2}}\right] dr'\right],
\end{equation}
so that the brane action in the monster case only differs from that of black hole by a term \emph{independent} of $r$. This means that in particular, the turning points, if any, of the actions are the same. 

For ordinary star, the term $1-(f\epsilon^{-1})^{1/2}$ is positive as we explained above, and so the brane action is in fact positive at the boundary. However, for the shell monster of which $\epsilon=\epsilon_0=\text{const.}$, we now see that even though the brane action is exactly the same as Eq.(\ref{starboundary}), the action can become negative and \emph{stays negative} if 
\begin{equation}\label{k=0}
 \sqrt{\epsilon} < \left(6\mathlarger{\int}_{r_{\text{eh}}}^R r'^2 f^{\frac{1}{2}}dr'\right)\left[\pi L^2 \Theta A + 2\left(R^3-r^3_{\text{eh}}\right)\right]^{-1} \equiv \sqrt{\epsilon_\infty}.
\end{equation}
If infinite reservoir of negative action does imply no stable solutions exist regardless of backreaction, then this allows us to slay all shell monsters spacetime whose Euclidean metric induces flat curvature on the conformal boundary, provided that Eq.(\ref{k=0}) holds. \emph{Mutatis mutandis}, the argument should work for all monsters and their kin whose metric asymptotes to that of flat black hole, \emph{as long as their ``problematic region'' does not lie behind the true event horizon}, since otherwise as we have already mentioned, the region will be excised upon Wick-rotating to Euclidean signature.

The question that we have to settle at this point is whether infinite supply of negative action does imply non-existence of stable solutions. The answer is probably \emph{no} in general. Consider for example the remarkable black hole solution found by D. Klemm, V. Moretti, and L. Vanzo \cite{KMV}. The KMV black holes can have spatially flat spatial sections, dubbed the KMV$_0$ solution \cite{KMV} and is in fact a kind of rotating (more appropriately, shearing) planar black hole. It is shown in \cite{McInnes0} that \emph{no matter how small} the angular momentum is, the brane action will eventually become negative and \emph{stay} negative. It is thus tempting to conclude that such solutions are not physical, since emission of finitely many brane-anti-brane pairs does not seem capable of getting rid of the infinite negative action. However, surprisingly, \emph{it is possible to do so}. One only needs to recall that angular momentum of black hole, like its mass, can be changed by physical processes. When calculating the brane action for the KMV$_0$ black hole, one uses the original stationary black hole metric, but once brane emission is triggered, one has to eventually take into account the backreaction to the black hole spacetime. Specifically, we can imagine that branes are nucleated in pairs with a nonzero total angular momentum, and they carry it away and thus reduce the angular momentum of the black hole. Since the latter is only \emph{finite}, and in fact could be very small, it is easy to imagine that such process could reduce it completely to zero. Thus, by emitting only finitely many brane-anti-brane pairs, \emph{the infinite supply of negative action can in principle, be removed}. Therefore, we should ask: is there any finite parameter in the monster case which, when changed by a finite amount, removes the infinite negative action?\footnote{Note that KMV$_0$ black hole is not spatially flat at the boundary of AdS, but only \emph{conformally} flat \cite{McInnes-1}, i.e., it does not have quite the same geometry as the monster under consideration. However all we want to stress here is that, in principle infinite amount of negative action of a given geometry could go away by tuning certain parameter by finite amount; this is regardless of the curvature at the boundary.}

Since the monster considered here is characterized by its mass, as well as the equation of states of the fluid, one obvious way out is to \emph{collapse into a black hole fast enough}, and in the process, pushes the problematic region pass the (true) horizon. Once the entire problematic region is behind the horizon, its Wick-rotated Euclidean metric will remove all problems. To recap, in order for us to claim that the monster has infinite supply of negative action and hence does not represent a physical solution, we need to establish that the infinite negative action cannot be removed by changing finite parameter. The corresponding parameter is the size of the problematic region, which is finite, and hence by collapsing pass the horizon, can remove the entire infinite supply of negative action in finite proper time. For our plan to work then, we must establish that the monsters do not in fact have enough time to collapse into black hole before Seiberg-Witten instability kicks in, thereby modifying the spacetime considerably to the effect that \emph{we can no longer trust that the negative action can be removed via gravitational collapse}. We will come back to this important issue later on. 

For now, let us consider the brane action and consider the point at which the action vanishes. This value of $r=r_{0}$ must satisfy the equation
\begin{equation}
r_{0}^2\left[-\frac{2\bar{M}}{r_{0}} + \frac{r_{0}^2}{L^2}\right]^{\frac{1}{2}} - \frac{3}{L}\int_{r_{\text{eh}}}^R r'^2\left(f\epsilon^{-1}\right)^{\frac{1}{2}} - \frac{3}{L}\int_R^{r_{0}} r'^2~ dr' = 0.
\end{equation}
This leads to
\begin{equation}
r_{0}^3\left[\left(-\frac{2\bar{M}L^2}{ r_{0}^3}+1\right)^{\frac{1}{2}}-1\right] + R^3 = 3\int_{r_{\text{eh}}}^{R} r'^2\left(f\epsilon^{-1}\right)^{\frac{1}{2}}~dr' > 0.
\end{equation}
Interestingly, we observe that since the right hand side is positive, with $\epsilon=\epsilon_0$ assumed independent of $r$, it cannot be arbitrary small since the left hand side is bounded away from upper bound $R^3$. However we have previously shown that for $\epsilon < \epsilon_\infty$, the action is negative at infinity. We thus have to consider two quantities: $\epsilon_\infty$ and $\epsilon_{\text{min}}$, the latter refers to the lower bound of $\epsilon$ that allows the action to vanish. Explicitly,

\begin{equation}
\sqrt{\epsilon_{\text{min}}}= \frac{3 \displaystyle \int_{r_{\text{eh}}}^R r'^2f^{\frac{1}{2}} dr'}{\text{max}\left[ r^3\left[(1-{\dfrac{2\bar{M}L^2}{r^3}})^{\frac{1}{2}}-1\right]\right]+R^3}
= \frac{3 \displaystyle \int_{r_{\text{eh}}}^R r'^2 f^{\frac{1}{2}} dr'}{-\bar{M}L^2 + R^3}
= \frac{6 \displaystyle \int_{r_{\text{eh}}}^R r'^2 f^{\frac{1}{2}} dr'}{2R^3 - r_{\text{eh}}^3},
\end{equation}
where we have used the fact that the event horizon is $r_{\text{eh}}=(2\bar{M}L^2)^{1/3}$.

On the other hand, we have from Eq.(\ref{k=0}), 
\begin{equation}
\sqrt{\epsilon_\infty} = \frac{6 \displaystyle \int_{r_{\text{eh}}}^R r'^2f^{\frac{1}{2}} dr'}{2R^3 - r_{\text{eh}}^3 + (\pi L^2 \Theta A - r_{\text{eh}}^3)}.
\end{equation}
Therefore, we can consider several cases. The first possibility is if the brane tension satisfies the inequality $\Theta > M/(2\pi^3 K^3)$. This corresponds to $\epsilon_\infty < \epsilon_{\text{min}}$. We then have three subcases:
\begin{itemize}
\item[(1.)] If $\epsilon \leq \epsilon_\infty < \epsilon_{\text{min}}$, then the brane action $S$ never crosses the $r$-axis, and remains negative at infinity, with the marginal case occurs when equality is attained. 
\item[(2.)] If $\epsilon_\infty < \epsilon < \epsilon_{\text{min}}$, then $S$ never crosses the $r$-axis, and is positive at infinity. This is the case for ordinary star (NEC preserving fluid) provided $0 < (f\epsilon^{-1}) < 1$. 
\item[(3.)] If $\epsilon_{\infty}<\epsilon_{\text{min}}<\epsilon$, then $S$ crosses zero, and remains positive at infinity.  
\end{itemize}

Likewise, for brane action satisfying $\Theta < M/(2\pi^3 K^3)$. we have $\epsilon_\infty > \epsilon_{\text{min}}$. We then also have three subcases:
\begin{itemize}
\item[(I.)] If $\epsilon_{\text{min}} < \epsilon < \epsilon_\infty$, then $S$ becomes zero at some finite value of $r$ but negative at infinity. This case cannot happen since the brane action $S$, like the brane action of the black hole counterpart, is monotonically increasing (and tends to a constant asymptotically) and thus has no turning point.
\item[(II.)] If $\epsilon < \epsilon_{\text{min}} < \epsilon_{\infty}$, then $S$ never becomes zero and remains negative at infinity.
\item[(III.)] If $\epsilon_{\text{min}} < \epsilon_{\infty} < \epsilon$, then $S$ is positive at infinity but crosses the $r$-axis at some finite value. 
\end{itemize} 

Going back to Eq.(\ref{shift}), we see that since the brane action for monsters only differs to that of black hole by a shift that only depends on $\epsilon$, which we sketched in Fig.1. It is thus possible for the action to have only \emph{finite} amount of negative reservoir, namely case (3) and case (III). We shall refer to such cases as ``small monsters'', while those with infinite supply of negative action, ``large monsters''. The instability of such monsters is thus of the Maldacena-Maoz type \cite{MaldacenaMaoz}. 

\begin{center}
\begin{figure}
\includegraphics[width=7.0 in]{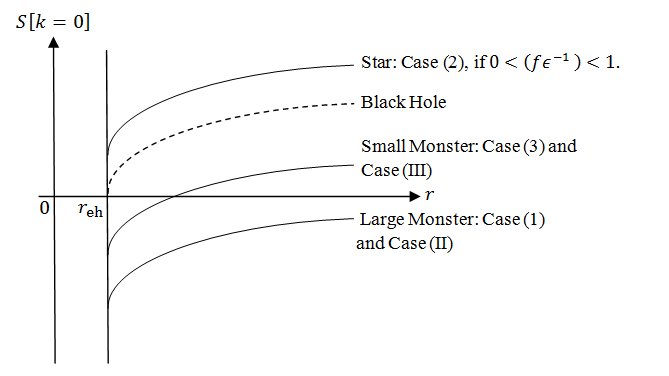}
\caption{The schematic graphs of brane actions for various geometries with flat spatial sections: star (NEC preserving fluid), black hole, small monster (finite supply of negative action) and large monster (infinite supply of negative action). Note that the various actions only differ by constant shift. The range $r < r_{\text{eh}}$ is removed upon Wick-rotation and hence the graphs only start at $r_{\text{eh}}$.}
\end{figure}
\end{center}

Let us move on to investigate the $k=\pm 1$ cases. We note that for $k=1$ case, the instability is \emph{always} of Maldacena-Maoz type, with only finite supply of negative action. Nevertheless as we send $\epsilon$ towards zero, the action becomes \emph{very} negative, indicating the solution becoming more unstable. To see the behavior of the brane action more explicitly, we note that for $k=1$ topological black hole, the brane action is
\begin{flalign}
S[\text{Black Hole}(k=1)](r)&=r^2f^{\frac{1}{2}} - \int_{r_{\text{eh}}}^r r'^2 dr'=r^2\left(1-\frac{2M}{r}+\frac{r^2}{L^2}\right)^{\frac{1}{2}}-\frac{r^3-r^3_{\text{eh}}}{L}\\
&=\frac{Lr}{2}-ML + \frac{r^3_{\text{eh}}}{L} + \mathcal{O}\left(r^{-1}\right),
\end{flalign}
so that the brane action at infinity behaves linearly in $r$. This is clearly divergent, and so we cannot conclude as in $k=0$ case that the action will become negative at the boundary even if $\epsilon_0$ is taken to be arbitrarily small. Indeed, at any \emph{fixed finite value} of $r=r^*$, the action $S[\text{Monster}(k=1)](r^*)$ will be negative if 
\begin{equation}\label{contradiction}
\sqrt{\epsilon} < \left(\frac{3}{L}\mathlarger{\int}_{r_{\text{eh}}}^R r'^2 f^{\frac{1}{2}}dr'\right)  \left(S[\text{Black Hole}](r^*) + \frac{3}{L}\mathlarger{\int}_{r_{\text{eh}}}^R r'^2 dr'\right)^{-1}.
\end{equation}
In particular, if we do take limit $r \to \infty$, the action will only be negative at the boundary if $\epsilon=\epsilon_0 < 0$. However by definition, $\epsilon >0$. In other words for the shell monster with $k=1$, the action at the boundary is actually \emph{positive}. It is however clear that for finite values of $r=r^*$, $\epsilon$ can be chosen small enough so that the action becomes negative at points $r_{\text{eh}} \leq r \leq r^*$. In other words, the instability of such monsters are of the Maldacena-Maoz type \cite{MaldacenaMaoz}. 

Finally, for $k=-1$ case, every brane action will be unbounded below as shown in Fig.2. We are now ready to interpret the results of our calculations.

\begin{center}
\begin{figure}
\includegraphics[width=7.0 in]{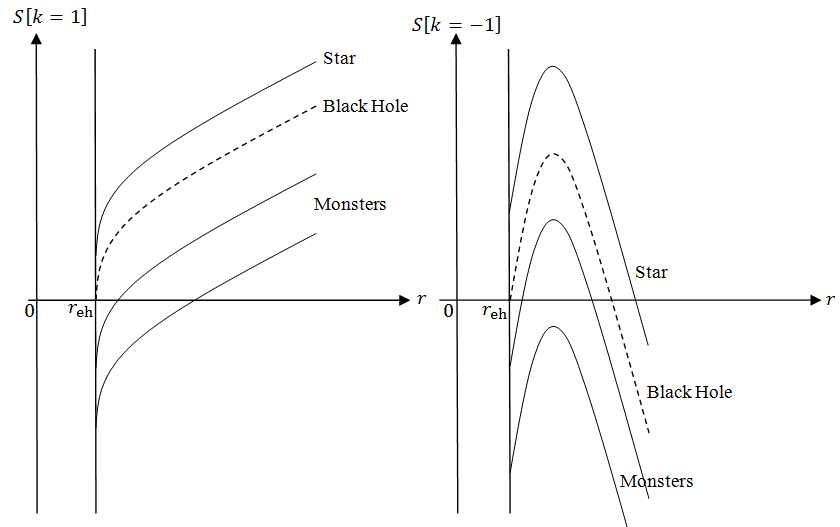}
\caption{The schematic graphs of brane actions for various geometries for positively curved spatial section and negatively curved spatial section. Note that for positively curved geometries, all monsters have only finite supply of negative action; while monsters in negatively curved background have infinite reservoir of negative action.  The range $r < r_{\text{eh}}$ is removed upon Wick-rotation and hence the graphs only start at $r_{\text{eh}}$.}
\end{figure}
\end{center}

\section{Collapsing vs. Destabilizing a Monster}

As already pointed out in \cite{McInnes0}, what is relevant about instability is the time scale it takes for the instability to manifest itself. If the instability time scale is very long, the physical system in question may change its properties due to other effects long before the instability can develope. In the case studied in \cite{McInnes0}, the planar KMV black hole is used to study quark gluon plasma formed in heavy ion collision. Since the latter quickly expands and cools after collision, any effect which takes longer than the hadronization time will simply not be observed. As a consequence, one does not expect to observe Seiberg-Witten instability at low angular momenta since the point $r_{\text{neg}}$, at which the brane action of planar KMV black holes becoming zero (and negative beyond the point), is further away from the horizon the slower the rotation is. 

The main idea for causality is this: If the brane nucleation happens far away from the black hole, the black hole cannot know about the branes moving around in the bulk until enough time has passed that the brane reaches the vicinity of the hole to significantly perturb and change its geometry, which consequently changes also the original action. To estimate this time scale, one can calculate the amount of time for the brane to free fall inwards to the horizon \cite{McInnes0}. This will be related to the distance the brane has to travel, and thus involves an integral $I$ over $r$ from the horizon $r_{\text{eh}}$ to the point at which the brane action vanishes $r_{\text{neg}}$. 

In our case, if a monster has collapsed into black hole before it knows about the brane nucleation, then Seiberg-Witten instability will simply \emph{not} be observed. What is the time scale for gravitational collapse to remove the problematic region beyond the horizon? This will involve free fall of particle from the edge of the fluid configuration at $r=R$. That is, the process involves the same integral $I$ over $r$, but now from the horizon $r_{\text{eh}}$ to $R$. For usual black hole cases in which the brane action only becomes negative at relatively large $r=r_\text{neg}$, we will then proceed by comparing whether $r_{\text{neg}} < R$ or the other way round to determine which effect is at work earlier, here for the monster case the situation is completely different: \emph{all monsters have negative action starting right at the horizon}! In other words the monster configurations are almost immediately aware of the brane nucleations in the close vicinity of the fluid surface (if not the horizon itself; since the horizon is still within the fluid when it is first formed) and backreaction is swift to begin changing the monster to some other stable configurations. Therefore the backreaction should take place at the same order of time scale, if not faster, than the time it takes for gravitational collapse. As a consequence we cannot be confident that monsters will always collapse into black hole (at least in AdS spacetime). If the instability is of Maldacena-Maoz type (all $k=1$ monsters and $k=0$ little monsters), we have hope to obtain a new (albeit unknown) final configuration after the brane emission removes all the supply of negative action. If there is an infinite reservoir of negative action ($k=0$ large monsters and $k=-1$ monsters), since we can no longer trust that gravitational collapse can remove the problematic region and there is no other parameters that can be changed by finite amount yet removing infinite supply of negative action, it seems plausible that these configurations are unphysical, i.e. not a solution to full theory of quantum gravity.


\section{Discussion}

We thus conclude that $k=-1$ monster configurations are completely unstable and $k=0$ ones are also completely unstable if $\epsilon_0$ violates the bound 
\begin{equation}\label{k=0 bound}
 \sqrt{\epsilon} \geq \left(6\mathlarger{\int}_{r_{\text{eh}}}^R r'^2 f^{\frac{1}{2}}dr'\right)\left[\pi L^2 \Theta A + 2\left(R^3-r^3_{\text{eh}}\right)\right]^{-1},
\end{equation}
in the sense that there is an infinite supply of negative action. Furthermore this instability is at work at the same time scale, if not earlier than gravitational collapse time scale, and thus the brane action cannot be trusted to be made positive by emission of finite number of brane-anti-brane pairs. This signals that such configurations are very likely not valid solutions of the full theory of quantum gravity. The competition between gravitational collapse and Seiberg-Witten instability will require further careful analysis to determine whether gravitational collapse can be the winner. In this work we only pointed out that it is not clear that monsters in anti-de Sitter spacetime will always evolve to become black holes. The case $k=1$ is somewhat different: the action is always positive at infinity. The finite reservoir of negative action can be reasonably removed via brane-anti-brane pair productions. We thus expect a new stable non-monster configuration at the end of the backreaction. In all cases, the final state of monsters, perhaps worthy of the name monsters, remains elusive, much like their mythological counterparts.

We now comment on the kin of monsters. We note that in our analysis via Seiberg-Witten instability, it is crucial that we perform Wick-rotation, a process that eliminates the interior of black hole event horizon. For monsters, which do not have true event horizon, their interior is not removed under Wick-rotation, and this is crucial in the analysis of some types of monsters, e.g. the concentric monsters that may have their problematic region positioned behind the horizon (the formal horizon identified by the exterior vacuum spacetime, hence not real horizon). This also implies that while our work may slay some classes of monsters and some of their kin, \emph{any kin of monster with problematic regions hiding behind a true event horizon cannot be ruled out by the this kind of analysis}. This includes, but not limited to, bag-of-gold spacetime with closed FRW universe hiding behind Einstein-Rosen bridge of a Schwarzschild black hole (more precisely, its AdS version), and maximally extended (AdS-)Schwarzschild black hole. Curiously it is not entirely clear whether these types of configurations with large volumes bounded by true event horizon is problematic in AdS/CFT correspondence. In \cite{Marolf}, it is raised that AdS/CFT seems to predict that quantum tunneling to create bag-of-gold spacetime is not possible, however it is also pointed out in the same work that there \emph{are} theories in which bag-of-gold spacetime \emph{is} allowed in AdS, and in fact various works have suggested that such theories are dual to a product theory \cite{Maldacena, inflation}, with one factor being the usual CFT and the other being some new set of degrees of freedom. Here, we do not claim to have shed any light on this interesting issue; we only argue that at least some types of monsters seem not to be plausible in quantum gravity. If this suggests that indeed there is difficulty to admit all monsters solutions in the final theory, then even if a bag-of-gold spacetime exists, the bag probably develops either together with, or after the black hole horizon formation, instead of passing through an intermediate monster stage. 

Finally we make some comments on the puzzles of black hole entropy for the sake of completeness. No less puzzling than information loss paradox is the problem of the origin of black hole entropy. Consider a star,  its entropy content scales like $S \sim A^{3/4}$ \cite{HR}, which is minuscule compared to entropy of a black hole $S=A/4$. This is consistent with the Second Law of Thermodynamics, i.e. entropy \emph{should} be expected to increase. However, the entropy gap between $A^{3/4}$ and $A$ leaves one wondering: where did the huge amount of entropy increase come from when a star collapses into a black hole? Perhaps a bag-of-gold does develop to conceal large amount of information behind the horizon. Before proceeding, we should stress that it is still an unsettled issue \cite{JMR} whether Bekenstein-Hawking entropy measures only the degrees of freedom on the horizon of the black hole, i.e. the surface area, or indeed measures all the degrees of freedom of a black hole, including its interior volume. These two interpretations are so-called ``weak form'' and ``strong form'' of Bekenstein-Hawking entropy interpretations, respectively \cite{MS, Smolin, Sorkin, Jacobson, SL}. If the weak form interpretation is correct, then there is no direct relation between the entropy of the black hole interior (i.e. the information it can store), and the mass of the black hole (which is related to its area). Thus, as with the case of bag-of-gold spacetime, a black hole of any fixed finite mass can contain in its interior, arbitrary high amount of information. See \cite{Marolf} for the same puzzle regarding whether the weak form or strong form is correct, albeit in the context of AdS/CFT correspondence. 

Note that if one believes in the weak form interpretation, then there is no information loss paradox to solve, since all the information can be safely stored in the black hole even if Hawking radiation is completely thermal and carries away no information, until finally the information comes out during the Planck scale, at which point effective field theory is expected to fail anyway, and so all bets are off. Furthermore, black holes evaporation may well end in a remnant instead of completely evaporates away. The remnants may in principle stores large amount of entropy in its bag-of-gold. The case for remnants was argued in e.g. \cite{ACS}. Note that in that work, the entropy of the remnant is shown to be ``small''. \emph{That entropy}, however, is just the Bekenstein-Hawking entropy, which, if one believes in the weak form interpretation, does not really measure the information content of the black hole. The argument that monsters lead to problems when considering eventual loss of information is therefore contingent to the assumption that the strong form interpretation is correct. Unfortunately our work does not shed any light on this age old problem regarding interpretation of Bekenstein-Hawking interpretation and its relation to the information loss paradox. 

\acknowledgments
Yen Chin Ong would like to thank Brett McInnes for helpful comments, Steve Hsu for various clarifications, Keisuke Izumi for useful discussions, and Siyun Zhou for fun chats over quantum physics. The authors would also like to thank Sabine Hossenfelder, for inspiring their investigation into monsters during a discussion at NORDITA during the 13th Marcel Grossmann Meeting, and for her much appreciated comments on this work. Pisin Chen is supported by Taiwan National Science Council under Project No. NSC 97-2112-M-002-026-MY3, by Taiwan's National Center for Theoretical Sciences (NCTS), and by US Department of Energy under Contract No. DE-AC03-76SF00515. Yen Chin Ong is supported by the Taiwan Scholarship from Taiwan's Ministry of Education.


\begin{thebibliography}{100}

\bibitem{HR} S. D. H. Hsu, D. Reeb, \emph{Black Hole Entropy, Curved Space and Monsters}, Phys. Lett. B \textbf{658} (2008) 244, \href{http://arxiv.org/abs/0706.3239v2}{[0706.3239v2 [hep-th]]}.

\bibitem{HR2} S. D. H. Hsu, D. Reeb, \emph{Monsters, Black Holes and the Statistical Mechanics of Gravity}, Mod. Phys. Lett. A \textbf{24} (2009) 1875, \href{http://arxiv.org/abs/0908.1265}{[0908.1265v1 [gr-qc]]}.

\bibitem{Wheeler} J. A. Wheeler, \emph{Relativity, Groups, and Fields}, edited by B. S. DeWitt and C. M. DeWitt, Gordon and Breach, New York (1964).

\bibitem{SWJ} R. D. Sorkin, R. M. Wald, Z. J. Zhang, \emph{Entropy of Self-Gravitating Radiation}, Gen. Rel. Grav. \textbf{13} (1981) 1127.

\bibitem{Wald} R. M. Wald, \emph{General Relativity}, University Of Chicago Press, First Edition (1984).

\bibitem{Ted} T. Jacobson, \emph{When Is $g_{tt}g_{rr}=-1$?}, Class. Quant. Grav. \textbf{24} (2007) 5717, \href{http://arxiv.org/abs/0707.3222v3}{[0707.3222v3 [gr-qc]]}. 

\bibitem{Stuchlik} Z. Stuchlik, \emph{Spherically Symmetric Static Configurations of Uniform Density in Spacetimes with a Non-Zero Cosmological Constant}, ActaPhys. Slov. \textbf{50} (2000) 219, \href{http://arxiv.org/abs/0803.2530v1}{[0803.2530v1 [gr-qc]]}.

\bibitem{ACS} R. J. Adler, P. Chen and D. I. Santiago, \emph{The Generalized Uncertainty Principle and Black Hole Remnants}, Gen. Rel. Grav. \textbf{33} (2001) 2101, \href{http://arxiv.org/abs/gr-qc/0106080}{[gr-qc/0106080v1]}.

\bibitem{Marolf} D. Marolf, \emph{Black Holes, AdS, and CFTs}, Gen. Rel. Grav. \textbf{41} (2009) 903, \href{http://arxiv.org/abs/0810.4886v2}{[0810.4886v2 [gr-qc]]}.

\bibitem{HR3} S. D. H. Hsu, D. Reeb, \emph{Unitarity and Hilbert Space of Quantum Gravity}, Class. Quant. Grav. \textbf{25} (2008) 235007, \href{http://arxiv.org/abs/0803.4212}{[0803.4212v2 [hep-th]]}.

\bibitem{McInnes} B. McInnes, \emph{Horizon Complementarity and Casimir Violations of the Null Energy Condition}, \href{http://arxiv.org/abs/0811.4465v2}{[0811.4465v2 [hep-th]]}.

\bibitem{NEC} R. V. Buniy, S. D. H. Hsu, B. M. Murray, \emph{The Null Energy Condition and Instability}, Phys. Rev. D \textbf{74} (2006) 063518, \href{http://arxiv.org/abs/hep-th/0606091}{[hep-th/0606091]}. 

\bibitem{FR} P. G. O.~Freund, M.~A.~Rubin, \emph{Dynamics of Dimensional Reduction}, Phys. Lett. B \textbf{97} (2) (1980) 233.

\bibitem{SeibergWitten} N. Seiberg and E. Witten, \emph{The D1/D5 System and Singular CFT}, JHEP \textbf{9904} (1999) 017, \href{http://arxiv.org/abs/hep-th/9903224}{[hep-th/9903224v3]}.

\bibitem{Kleban} M. Kleban, M. Porrati, R. Rabadan, \emph{Stability in Asymptotically AdS Spaces}, JHEP \textbf{0508} (2005) 016, \href{http://arxiv.org/abs/hep-th/0409242}{[hep-th/0409242v1]}.

\bibitem{Schwinger} J. Schwinger, \emph{On Gauge Invariance and Vacuum Polarization}, Phys. Rev. \textbf{82} (1951) 664. 

\bibitem{Barbon} Jos\'e L.F. Barb\'on, Javier Mart\'inez-Mag\'an, \emph{Spontaneous Fragmentation of Topological Black Holes}, JHEP \textbf{08} (2010) 031, \href{http://arxiv.org/abs/1005.4439}{[1005.4439v1 [hep-th]]}.

\bibitem{McInnes5} B. McInnes, \emph{Topologically Induced Instability in String Theory}, JHEP \textbf{0103} (2001) 031, \href{http://arxiv.org/abs/hep-th/0101136v2}{[hep-th/0101136v2]}.

\bibitem{BKLS} I. Bredberg, C. Keeler, V. Lysov, A. Strominger, \emph{Cargese Lectures on the Kerr/CFT Correspondence}, Nucl. Phys. B, Proc. Suppl. 216 (2011) 194, \href{http://arxiv.org/abs/1103.2355}{[[1103.2355v3 [hep-th]]}.

\bibitem{MaldacenaMaoz} Juan Maldacena, Liat Maoz, \emph{Wormholes in AdS}, JHEP \textbf{0402} (2004) 053, \href{http://arxiv.org/abs/hep-th/0401024}{[hep-th/0401024v2]}.


\bibitem{McInnes2} B. McInnes, \emph{Black Hole Final State Conspiracies}, Nucl. Phys. B \textbf{807} (2009) 33, \href{http://arxiv.org/abs/0806.3818}{[0806.3818v2 [hep-th]]}.

\bibitem{SmithMann} W. L. Smith, R. B. Mann, \emph{Formation of Topological Black Holes from Gravitational Collapse}, Phys. Rev. D \textbf{56} (1997) 4942, \href{http://arxiv.org/abs/gr-qc/9703007}{[gr-qc/9703007]}.

\bibitem{KMV} D. Klemm, V. Moretti, L. Vanzo, \emph{Rotating Topological Black Holes}, Phys. Rev. D \textbf{57} (1998) 6127; Erratum-ibid. \textbf{60} (1999) 109902, \href{http://arxiv.org/abs/gr-qc/9710123}{[gr-qc/9710123]}.

\bibitem{McInnes0} B. McInnes, \emph{Fragile Black Holes and an Angular Momentum Cutoff in Peripheral Heavy Ion Collisions}, Nucl. Phys. B \textbf{861} (2012) 236, \href{http://arxiv.org/abs/1201.6443v2}{[1201.6443 [hep-th]]}.

\bibitem{McInnes-1} B. McInnes, \emph{Shearing Black Holes and Scans of the Quark Matter Phase Diagram}, \href{http://arxiv.org/abs/1211.6835}{[1211.6835v2 [hep-th]]}.
 
\bibitem{Maldacena} J. M. Maldacena,  \emph{Eternal Black Holes in AdS}, JHEP \textbf{0304} (2003) 021, \href{http://arxiv.org/abs/hep-th/0106112}{[hep-th/0106112v6]}.

\bibitem{inflation} B. Freivogel, V. E. Hubeny, A. Maloney, R. C. Myers, M. Rangamani, S. Shenker, \emph{Inflation in AdS/CFT}, JHEP \textbf{0603} (2006) 007, \href{http://arxiv.org/abs/hep-th/0510046}{[hep-th/0510046v4]}.

\bibitem{JMR} T. Jacobson, D. Marolf, C. Rovelli, \emph{Black Hole Entropy: Inside or Out?}, Int. J. Theor. Phys. \textbf{44} (2005) 1807, \href{http://arxiv.org/abs/hep-th/0501103}{[hep-th/0501103v1]}.

\bibitem{MS} F. Markopoulou, L. Smolin, \emph{Holography in a Quantum Spacetime}, \href{http://arxiv.org/abs/hep-th/9910146}{[hep-th/9910146v1]}.

\bibitem{Smolin} L. Smolin, \emph{The Strong and Weak Holographic Principles}, Nucl. Phys. B \textbf{601} (2001) 209, \href{http://arxiv.org/abs/hep-th/0003056}{[hep-th/0003056v1]}.

\bibitem{Sorkin} R. D. Sorkin, \emph{The Statistical Mechanics of Black Hole Thermodynamics}, ``Black Holes and Relativistic Stars'', p.177, edited by R. M. Wald, University of Chicago Press (1998), \href{http://arxiv.org/abs/gr-qc/9705006}{[gr-qc/9705006v2]}..

\bibitem{Jacobson} T. Jacobson, \emph{On the Nature of Black Hole Entropy},  AIP Conf. Proc. \textbf{493} (1999) 85, \href{http://arxiv.org/abs/gr-qc/9908031}{[gr-qc/9908031]}.

\bibitem{SL} S. Hossenfelder, L. Smolin, \emph{Conservative Solutions to the Black Hole Information Paradox}, Phys. Rev. D \textbf{81} (2010) 064009, \href{http://arxiv.org/abs/0901.3156v1}{[0901.3156 [gr-qc]]}.


\end{thebibliography}
\end{document}